\documentstyle[aps,epsf]{revtex}

\draft

\begin{document}
\title{Phenomena of spin rotation and oscillation of particles (atoms, molecules)
containing in a trap blowing on by wind of high energy particles in storage ring - 
new method of measuring of spin-dependent part of zero-angle coherent scattering amplitude}
\author{ V. G. Baryshevsky\thanks{%
E-mail bar@inp.minsk.by}}
\address{Institute of Nuclear Problems, Belarusian State University \\
St. Bobryiskaya 11, 220050, Minsk, Republic of Belarus }
\date{\today}
\maketitle

\begin{abstract}
New experiment arrangement to study the
spin rotation and oscillation of particles of a gas target, through which a beam of 
high energy particles passes, is discussed. Such experiment arrangement make it 
realizable for a storage ring and allows one to study the zero-angle scattering amplitude
at the highest possible energies. Life-time of a particle beam in a storage ring can 
reach several hours and even days. Life-time of a particle in a gas target (gas 
trap) is long too. Particles circulate in the storage ring with the frequency $\nu$ of 
several MHz. This yields to the $\nu$-fold increase of the density of the beam of high energy 
particles, which blows a gas trap, in comparison with the single-pass case. 
Finally, this causes the perfectly acceptable angle of 
the spin rotation of particles containing in the gas cell.
Relation between the index of refraction and effective potential energy of 
a particle in a medium is discussed. Phenomenon of the spin rotation of a particle
captured to a trap under the action of a beam of polarized particles is 
considered. Expressions for effective potential 
energy and angle of spin rotation are derived for particles in a trap.
Rotation and oscillation of deuteron spin is studied. 
Estimations for angle of rotation show
that the effect can be experimentally observed.
\end{abstract}



\section{Introduction}

Investigation of spin-dependent interactions of elementary particles
at high energies is a very important part of a program of scientific research
has been preparing for carry out at storage rings (RHIC, CERN, COSY). It is well 
known in experimental particle physics how to measure  a total spin-dependent 
cross-section of proton-proton (pp) and proton-deuteron (pd) 
(or proton-nucleus (pN))
and deuteron-nucleus (dN)interactions.

Through analicity we can get dispersion relations between the real and imaginary parts 
of the forward scattering amplitude. These relations are very valuable for analyzing 
interactions, especially if we know both real and imaginary parts of the forward scattering 
amplitude in a broad energy range through the independent experimental measurements.

There are several experimental possibilities for the indirect measurement of the real part 
of the forward scattering amplitude \cite{Lehar}.

{\bf Since no scattering experiment is possible in the forward direction, the determination of the 
real part of the forward amplitudes has always consisted in the measurement of well chosen elastic 
scattering observables at small angles and then in the extrapolation of these observables towards 
zero angle \cite{Lehar}}. All of these methods, however, contain discrete ambiguities in the 
reconstruction of the forward scattering matrix, which can be removed only by new independent
measurements. Consequently,what is needed is a direct reconstruction of the real part of the forward 
scattering matrix such we have in the case of the imaginary part through the measurement of a total 
cross section.

It has been shown in [2-9] 
that there is an unambiguous method which makes possible the direct measurement
of the real part of the spin-dependent forward scattering amplitude in the high energy range.
This technique is based on the effect of proton (deuteron) beam spin rotation in a polarized nuclear target
and (it is very important)on the phenomenon of deuteron spin rotation and oscillation in a nonpolarized 
target. This technique uses the measurement of angle of spin rotation of high energy proton (neutron) in
conditions of transmission experiment - the so-called spin rotation experiment.

The analogous phenomenon for the thermal neutrons was theoretically predicted by Baryshevsky and Podgoretsky
in \cite{1rot} and experimentally observed by Abragam and Forte groups [11-13]
(the phenomena of nuclear precession of the neutron spin in a nuclear pseudomagnetic field of a target).

\begin{figure}[htbp]
\epsfxsize = 8 cm \centerline{\epsfbox{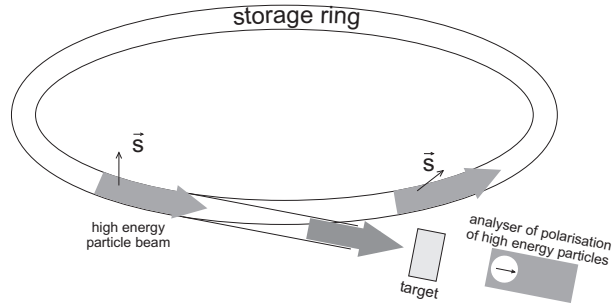}}
\caption{Conventional spin rotation experiment}
\label{conventional}
\end{figure}

It should be noticed that usually the spin state of high energy particles passing through a target is analyzed 
in the arrangement studying the spin rotation (see Fig.\ref{conventional}).

\begin{figure}[htbp]
\epsfxsize = 8 cm \centerline{\epsfbox{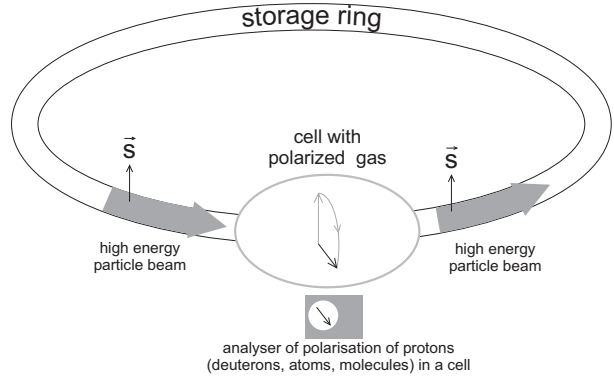}}
\caption{Measurement of spin rotation for particles in a gas cell}
\label{second}
\end{figure}

In the present paper it is shown that we can reverse experiment arrangement to study 
spin rotation and oscillation of particles of a gas target, through which a beam of 
high energy particles passes (see Fig.\ref{second}). 
Such experiment arrangement make it realizable for a 
storage ring and allows one to study the zero-angle scattering amplitude at the highest possible 
energies. Life-time of a particle beam in a storage ring can reach several hours. 
Life-time of a particle in a gas target (gas trap) is long too. Particles circulate 
in the storage ring with the frequency $\nu$ of several MHz. This yields to the $\nu$-fold 
increase of the current of the beam of high energy particle, which blows the gas trap, in comparison 
with the single-pass case. 
Finally, this causes the the perfectly acceptable angle of 
spin rotation of particles containing in the gas cell.

This paper, then, is organized as follows.
In section II the relation between the index of refraction and effective potential energy of 
a particle in medium is discussed. Section III considers phenomenon of spin rotation of a particle
captured to a trap under the action of a beam of polarized particles. Expressions for effective potential 
energy and the angle of spin rotation are derived for particles in the trap.
Section IV studies rotation and oscillation of deuteron spin. 
Estimations for the angle of rotation show
that the effect can be experimentally observed.

\section{The index of refraction and effective potential energy of particles in medium.} 

Close connection between the coherent elastic scattering amplitude at zero angle $f(0)$
and the refraction index of medium $N$ has been established 
as a result of numerous studies (see, for example, \cite{12, Goldberger}):
\begin{equation}
N=1+\frac{2\pi \rho }{k^{2}}f\left( 0\right)   
\label{refr_ind}
\end{equation}
where $\rho $ is the number of particles per $cm^{3}$, $k$ is the wave
number of a particle incident on a target. 

During derivation of (\ref{refr_ind}) it is supposed that $N-1 \ll 1$. 
If $k \rightarrow 0$ then $(N-1)$ grows and expression 
for $N$ has the form
\[
N^2=1+\frac{4\pi \rho }{k^{2}}f\left( 0\right)   
\label{refr_ind2}
\]

Let us consider refraction on the vacuum-medium boundary.

The wave number of the particle in vacuum is denoted $k$, 
$k^{\prime} = k N$ is the wave number of the particle in medium.

\begin{figure}[htbp]
\epsfxsize = 5 cm \centerline{\epsfbox{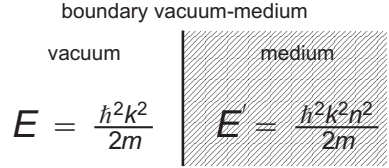}}
\caption{Kinetic energy of a particle in vacuum is not equal to that in medium.}
\end{figure}

As it can be seen, the kinetic energy of a particle in vacuum 
$E = \frac{\hbar^2 k^2}{2 m}$ is not equal to that in medium
$E^{\prime} = \frac{\hbar^2 k^2 N^2}{2 m}$.

From the energy conservation 
condition we immediately obtain the necessity to suppose that a particle  
in medium possesses effective potential energy
(see detailed theory in \cite{Goldberger}.) 
This energy can be found easily from 
evident equality
\[
E=E^{\prime}+U
\]
i.e.
\begin{equation}
U=E-E^{\prime }=- \frac{2 \pi {\hbar}^2}{m} {\rho} f(0).
\label{U1}
\end{equation}

Above we considered the rest target. But in storage rings moving bunches are 
usually used as a target. Therefore we should generalize expressions 
(\ref{refr_ind},\ref{U1}) for this case. Thus, let us consider cross-collision 
of two bunches of particles. Suppose that in the rest frame of storage ring
the particles of the first beam have the energy
$E_1$ and Lorentz-factor $\gamma_1$, whereas particles of the second beam are
characterized by the energy $E_2$ and Lorentz-factor $\gamma_2$. Let us recollect 
that phase of the wave in medium is Lorentz-invariant. Therefore, we can find it
by the following way. Let us choose the reference frame, where the second beam rests. 
As in this frame the substance of the second beam rests, then refraction index can be 
expressed in conventional form (\ref{refr_ind}):  
\begin{equation}
N_1^{\prime }=1+\frac{2\pi \rho_2^{\prime } }{{k_1^{\prime }}^{2}}f\left(E_1^{\prime}, 0\right),  
\label{refr_ind3}
\end{equation}
where $\rho_2^{\prime }=\gamma_2^{-1} \rho_2$ is the density of the bunch 2 in its rest
frame and $\rho_2$ is the density of the second bunch in the storage ring frame, 
$k_1^{\prime}$ is the wavenumber of particles of the first bunch in the rest frame of the bunch 2. 
Let us denote the length of the bunch 2 in its rest frame as $L$,
$L=\gamma_2 ~ l$, where $l$ is the length of this bunch in the storage ring frame.

Now we can find the change of the phase of the wave caused by the interaction of particle 1 with
the particles of bunch 2:
\begin{equation}
\phi=k_1^{\prime}(N_1^{\prime}-1)L=\frac{2 \pi \rho_2^{\prime}}{k_1^{\prime}} f(E_1^{\prime},0)~L
=\frac{2 \pi \rho_2}{k_1^{\prime}}{f(E_1^{\prime},0)}{k_1^{\prime}}~l  ,  
\label{phase}
\end{equation}

It is known that the ratio $\frac{f(E_1^{\prime},0)}{k_1^{\prime}}$ is invariant, so we can
write
$\frac{f(E_1^{\prime},0)}{k_1^{\prime}}=\frac{f(E_1,0)}{k_1}$, where $f(E,0)$ is the 
amplitude of elastic coherent forward scattering of particle 1 by the moving particle 2  
in the rest frame of the storage ring.

As a result
\begin{equation}
\phi=\frac{2 \pi \rho_2}{k_1} f(E_1,0)~l=\frac{2 \pi \rho_2}{k_1} f(E_1,0)~v_{rel}~t,  
\label{phase1}
\end{equation}
where $v_{rel}$ is the velocity of relative motion of the particle 1 and bunch 2,
if the first particle is nonrelativistic whereas the second that is relativistic,
then $v_{rel}\approx c$, where $c$ is the light velocity, $t$ is the time of interaction 
of the particle 1 with the bunch 2 in the rest frame of the storage ring.

The particle with velocity $v_1=\frac{\hbar k_1 c^2}{E_1}$ passes the
distance $z=v_1~t$ over time $t$. It should be noted that the path length $z$ differs from the 
length of bunch 2, because it moves.
Expression (\ref{phase1})can be rewritten by the use of $z$:
\begin{equation}
\phi=\frac{2 \pi \rho_2}{k_1} f(E_1,0)~\frac{v_{rel}}{v_1}~z=k_1 (N_1-1)z,  
\label{phase_z}
\end{equation}
where the index of refraction of the particle 1 by the beam of moving particles 2.
\begin{equation}
N_1=1+\frac{2 \pi {\rho}_2}{{k_1}^2}~\frac{v_{rel}}{v_1}f(E,0)
\label{r_ind_z}
\end{equation}
If $v_2=0$, the conventional expression (1) follows from (\ref{r_ind_z})

So, let us consider particles captured to a trap. They are nonrelativistic.
From (\ref{U1}) $U$ can be expressed as:
\begin{eqnarray}
U=- \frac{2 \pi {\hbar}^2}{m_1} {{\rho}_2} \frac{c}{v_1}f(E_1,0)= \\ \nonumber
=-2 \pi {\hbar} {\rho}_2 c  \frac{f(E_1,0)}{k_1}=\\ \nonumber
=-2 \pi {\hbar} {\rho}_2 c \frac{f(E_1^{\prime},0)}{k_1^{\prime}}=\\ \nonumber
=-\frac{2 \pi {\hbar}^2}{m_1 \gamma_2} {\rho}_2 f(E_1^\prime,0).
\label{U_}
\end{eqnarray}
where $\gamma_2$ is the Lorentz-factor of the bunch 2, the Lorentz-factor of the particle 1 is $\gamma_1=1$
$\hbar k_1^{\prime}=m_1 {c} \gamma_2$.

It should be noted that the amplitude of coherent elastic scattering at zero angle 
depends on T-matrix as follows:
\begin{equation}
f(E_1^{\prime},0)= - \frac{(2 \pi)^2 \hbar k_1^{\prime}}{v_1+v_2}T(E_1^\prime)=
(2 \pi)^2 m_1 \gamma_2 T(E_1^\prime)
\label{T-matrix}
\end{equation} 
i.e. the amplitude of forward scattering  is proportional  to the Lorentz-factor 
$\gamma_2$.
As a result, the quantity $\gamma_2^{-1} f(E_1^{\prime},0$ depends on 
{the energy of a particle} 
only due to possible dependence of T-matrix on energy.
From (\ref{U_},\ref{T-matrix}) one can obtain
\begin{equation}
U=(2 \pi)^3 \rho_2 T(E_1^\prime)
\label{T-matrix}
\end{equation}

\section{Rotation of spin of particles contained in a gas cell.}

In 1964 it was shown \cite{1rot}
that while slow neutrons are propagating through the target with polarized
nuclei, a new effect of nucleon spin precession occurred. It is stipulated by
the fact that neutrons in a polarized target possess two
refraction indices: $N_{\uparrow \uparrow }$ for neutrons with the spin
parallel to the target polarization vector and $N_{\uparrow \downarrow }$
for neutrons with the opposite spin orientation , $N_{\uparrow \uparrow
}\neq N_{\uparrow \downarrow }$. 
From the above it follows that in a medium with polarized nuclei, neuterons
possess two effective potential energies:
$U_{\uparrow \uparrow }$ for neutrons with the spin
parallel to the target polarization vector and $U_{\uparrow \downarrow }$
for neutrons with the opposite spin orientation , 
$U_{\uparrow \uparrow
}\neq U_{\uparrow \downarrow }$. This resembles interaction of neutrons
with a magnetic field, where the energy of interaction depends on the relative orientation of
neuteron spin and magnetic field.
According to \cite{1rot}, this comparison enables one to conclude that in a
target with polarized nuclei there is a nuclear pseudomagnetic field and the
interaction of an incident neutron with this field results in neutron spin
rotation. The results \cite{1rot}, had initiated experiments, which
proved the existence of this effect [11-13].

Let us consider the experiment arrangement shown in Fig.2. Gas of polarized particles 
(nuclei, atoms) is 
contained in a cell. A beam of relativistic polarized particles passes through the cell.
The relativistic beam is a moving target.
Let us study the polarization vector of particles captured to the cell.
To describe this vector behavior, it is necessary to find effective potential energy of interaction
of pacticles captured to the cell with the relativistic beam passing through it. First fo all
the amplitude of elastic coherent zero-angle
scattering of a nucleon by a polarized nucleon (nucleus) should be considered.

General form of this amplitude with allowance for strong electromagnetic
and weak interactions is given in \cite{2rot}. 
Below we shall consider 
the cell containing protons (nuclei with spin $1/2$). The beam of polarized
protons passes through it. Let us take into account only strong P-,T-invariant interactions.
In this case, the explicit structure of the elastic scattering amplitude \
of a particle with the spin 1/2 by a particle with the spin 1/2 \ (see, for example, 
\cite{14}) proceeds from the following simple 
reasoning. In our case,
the elastic scattering amplitude at zero angle depends on the spin operators
\ $\frac{1}{2}\overrightarrow{\sigma_b}$, $\frac{1}{2}\overrightarrow{\sigma _{c}}$ of an 
incident particle (beam) \
and that of a cell, and also on the relative momentum of particles $\overrightarrow{k}$,
i.e., on $\overrightarrow{n}=\frac{\overrightarrow{k}}{k}$ . Operators $%
\overrightarrow{\sigma_b},$ $\overrightarrow{\sigma _{c}}$ can be contained
in the expression for the amplitude only in the first degree, as higher
degrees \ of $\overrightarrow{\sigma }$ reduce either to a number or $%
\overrightarrow{\sigma }$. The combinations $\overrightarrow{\sigma_b},$ $%
\overrightarrow{\sigma _{c}}$ and $\overrightarrow{n}$ must be such that the
scattering amplitude is \ a scalar and invariant in space and time
reflections. These conditions definitely determine its general form:

\begin{equation}
\hat{F}=A+A_{1}\left( \overrightarrow{\sigma_b}\cdot \overrightarrow{%
\sigma _{c}}\right) +A_{2}\left( \overrightarrow{\sigma_b}\cdot 
\overrightarrow{n}\right) \left( \overrightarrow{\sigma _{c}}\cdot 
\overrightarrow{n}\right) .  \label{amplitude}
\end{equation}
By averaging the amplitude $\hat{F}$\ with the help of a spin matrix of
the density of particles $\rho _{b}$, the elastic coherent scattering
amplitude may be written as:

\begin{equation}
f=Sp\,\rho _{b}\,\hat{F}=A+A_{1}\left( \overrightarrow{\sigma_c}\cdot 
\overrightarrow{p}\right) +A_{2}\left( \overrightarrow{\sigma_c}\cdot 
\overrightarrow{n}\right) \left( \overrightarrow{n}\cdot \overrightarrow{p}%
\right)  \label{aver_amp}
\end{equation}
where $\overrightarrow{p}=Sp\,\rho _{b}$\thinspace $\overrightarrow{\sigma
_{b}}$ is the polarization vector of a particle in beam.

Amplitude $f$ can be expressed as
\begin{equation}
f=A+\overrightarrow{\sigma_c}\cdot \overrightarrow{g}  \label{ampl_g}
\end{equation}
where $\overrightarrow{g}=A_{1}\overrightarrow{p}+A_{2}\overrightarrow{n}%
\left(\overrightarrow{n}\cdot \overrightarrow{p}\right) $.

To simplify further reasoning let us consider the situation when vector $%
\overrightarrow{p}$ is either parallel to $\overrightarrow{n}$ \ ($%
\overrightarrow{p}\parallel \overrightarrow{n}$) or perpendicular to $%
\overrightarrow{n}$ ($\overrightarrow{p}\perp \overrightarrow{n}$).

In this case one has that $g\left( \overrightarrow{p}\parallel 
\overrightarrow{n}\right) =\left( A_{1}+A_{2}\right) \overrightarrow{p}$ and 
$g\left( \overrightarrow{p}\perp \overrightarrow{n}\right) =A_{1}%
\overrightarrow{p}$. Thus in these cases vector $\overrightarrow{g}$ is
directed along $\overrightarrow{p}.$ Selecting quantization axes parallel to 
$\overrightarrow{p}$, one can see that scattering amplitude $f_{\uparrow
\uparrow }=A+g$ \ of proton (nucleus) with spin parallel to $\overrightarrow{p}$ is
not equal to scattering amplitude $f_{\uparrow \downarrow }=A-g$ \ of
proton (nucleus) with spin antiparallel to $\overrightarrow{p}$. 
Hence,
corresponding 
effective potential energies 
are not equal to each other (i.e. $%
U_{\uparrow \uparrow }\neq U_{\uparrow \downarrow }$).

Let us consider a particle, which spin is parallel to the vector 
of polarization of beam particles $\overrightarrow{p}$.
From (8) it follows that effective potential energy of 
such a particle
captured in a trap is

\begin{equation}
U_{\uparrow \uparrow }=-\frac{2 \pi \hbar^2}{M ~\gamma_b} \rho_b f_{\uparrow \uparrow}(E^{\prime}_c,0).
\end{equation}
Here and below $M$ denotes 
the mass of the particle (nuclei) captured in the trap,
$\rho_b$ is the density of the beam rotating in a storage ring,
$f_{\uparrow \uparrow} (E^{\prime}_c,0)$ is the amplitude of elastic coherent forward scattering
of the particle captured in the trap (particle 1) by the particle of the beam (particle 2)
in the rest frame of the beam (particle 2), $E^{\prime}_c$ is the energy of the entrapped 
particle in the rest frame of the beam, $\gamma_b$ is the Lorentz-factor of the particle 2.
The latter frame is the laboratory 
frame by definition.
Effective potential energy of particle with opposite spin direction is

\begin{equation}
U_{\uparrow \downarrow }=-\frac{2 \pi \hbar^2}{M ~\gamma_b} \rho_b f_{\uparrow \downarrow}(E^{\prime}_c,0).
\end{equation}

Suppose that spin of the nucleon captured in a trap is
oriented at the certain angle to the vector $\overrightarrow{p}$. This state
of nucleon can be described as superposition of two states with
spins directed along and opposite to the vector $\overrightarrow{p}$. 
Spin wave function of nucleon at time moment $t=0$ can be expressed as:

\begin{equation}
\psi \left(0 \right) =\chi _{n}=\left( 
\pmatrix{
c_{1}\cr
c_{2}
}
\right)
\end{equation}

or

\begin{equation}
\psi \left( 0\right) =
c_{1}
\left( 
\pmatrix{
1 \cr
0
}
\right) +c_{2} \left( 
\pmatrix{
0 \cr 
1
}
\right).
\end{equation}

\noindent
The state $
\left( 
\pmatrix{
1 \cr
0
}
\right) $ possesses effective potential energy $U_{\uparrow \uparrow }$ and the state $%
\left( 
\pmatrix{
0 \cr
1
}
\right) $ is characterized by $U_{\uparrow \downarrow }$. 
Hence, the spin wave
function of a nucleon blowing on by a beam of high energy particles changes with time $t$ as
follows:

\begin{eqnarray}
\psi \left(t \right) &=&
c_{1}\,\,e^{-\frac{i}{\hbar}U_{\uparrow \uparrow }t}\left( 
\pmatrix{
1 \cr 
0
}
\right) +c_{2}\,\,e^{-\frac{i}{\hbar}U_{\uparrow \downarrow }t}\left( 
\pmatrix{
0 \cr
1
}
\right) = \\
& = & 
c_{1}{\psi}_{\uparrow \uparrow }(t)
\left( 
\pmatrix{
1 \cr
0
}
\right) +
c_{2}{\psi}_{\uparrow \downarrow }(t)
\left( 
\pmatrix{
0 \cr
1
}
\right).  \nonumber
\label{psi}
\end{eqnarray}

One can find nucleon polarization vector at the time moment $t$

\begin{equation}
\vec{p}_{n}(t)=\langle \psi |\vec{\sigma}|\psi \rangle~\langle \psi |\psi \rangle^{-1}.
\end{equation}
and as a result

\begin{eqnarray}
& &p_{nx}=2{Re}c_{1}^{\ast }c_{2}\psi _{\uparrow \uparrow }^{\ast }\psi _{\uparrow \downarrow }\langle \psi
|\psi \rangle ^{-1}, \nonumber \\
& &p_{ny}=2{Im}c_{1}^{\ast }c_{2}\psi _{\uparrow \uparrow }^{\ast
}\psi _{\uparrow \downarrow }\langle \psi |\psi \rangle ^{-1},  \label{p-project}  \\
& &p_{nz}=\left( \left| c_{1}\psi _{\uparrow \uparrow }\right| ^{2}-\left| c_{2}\psi
_{\uparrow \downarrow}\right| ^{2}\right) \langle \psi |\psi \rangle ^{-1}. \nonumber 
\end{eqnarray}

Suppose that nucleon spin in time moment $t=0$ is perpendicular 
to the beam polarization
vector $\overrightarrow{p}$. 
The axes $z$ is directed along the beam polarization
vector $\overrightarrow{p}$.
Direction of nucleon spin in time 
moment $t=0$ defines the axes $x$. In
this case $c_{1}=c_{2}=1/\sqrt{2}.$ Using (\ref{p-project}) we obtain

\begin{eqnarray}
& &p_{nx}=\cos \left( \Omega t \right) e^{- \frac{1}{2} \rho c \left( \sigma_{\uparrow \uparrow
}-\sigma_{\uparrow \downarrow }\right) t}\langle \psi |\psi \rangle ^{-1}, 
\nonumber \\
& &p_{ny}=\sin \left( \Omega t \right) e^{- \frac{1}{2} \rho c \left( \sigma_{\uparrow \uparrow
}-\sigma_{\uparrow \downarrow }\right) t\,}\langle \psi |\psi \rangle ^{-1},
\label{p-2} \\
& &p_{nz}=\frac{1}{2}\left(
e^{-\rho c \sigma_{\uparrow \uparrow} t}-
e^{-\rho c \sigma_{\uparrow \downarrow} t}
\right) \langle \psi |\psi
\rangle ^{-1}= \nonumber \\
& &=\frac
{\,
e^{-\rho c \sigma_{\uparrow \uparrow} t}-
e^{-\rho c \sigma_{\uparrow \downarrow} t}
}
{
e^{-\rho c \sigma_{\uparrow \uparrow} t}+
e^{-\rho c \sigma_{\uparrow \downarrow}t
}}.  \nonumber
\end{eqnarray}
$\langle \psi |\psi \rangle =\frac{1}{2}\left( \,
e^{-\rho c \sigma_{\uparrow \uparrow} t\,}+
e^{-\rho c \sigma_{\uparrow \uparrow} t\,}\right) \, ,~~
\Omega=\frac{Re (U_{\uparrow \uparrow}-U_{\uparrow \downarrow})}{\hbar}$
is the frequency of nucleon spin precession around polarization vector $\overrightarrow{p}$,
$\sigma_{\uparrow \uparrow}~(\sigma_{\uparrow \downarrow})$ is the total cross-section of
scattering in laboratory frame for particles with parallel (antiparallel) spins.
It should be noted that $Im~f(0)=\frac{k}{4~\pi}~\sigma$.

According to (\ref{p-2}) the polarization vector
of a nucleon, being blown by the beam of relativistic particles, 
rotates around nuclei polarization vector 
$\overrightarrow{p}$ at the angle

\begin{equation}
\vartheta =\Omega t=-\frac{2\pi \hbar}{M \gamma_b} \rho_b {Re}\left( f_{\uparrow \uparrow
}-f_{\uparrow \downarrow }\right) t  \label{theta}
\end{equation}

This rotation is similar to the spin rotation appearing in magnetic
field. Then we can conclude that 
the beam of relativistic particles
acts on spin of particles captured in a trap by the
nuclear pseudomagnetic field \cite{1rot}, which causes precession of spin of entrapped particles.

Above we considered only cases with $\overrightarrow{k}\perp \overrightarrow{p}$ and 
$\overrightarrow{k}\parallel \overrightarrow{p}$. In general case forward 
scattering amplitude is expressed by (\ref{ampl_g}): 
\[
\hat{f}(0)=A+\overrightarrow{\sigma }\cdot \overrightarrow{g}  
\]
Effective potential energy can be written as:
\[
\hat{U}=-\frac{2 \pi \hbar^2}{M \gamma_b} \rho_b \hat{f}(0).  
\]
The Shr\"{o}dinger equation for spin wave function of a particle
can be written as
\begin{equation}
i \hbar \frac{\partial \psi }{\partial t}=\hat{U} \psi.
\label{Shr_eq}
\end{equation}
and let to find behavior of spin of particles captured in a trap
at arbitrary orientation of $\overrightarrow{p}$ and $\overrightarrow{k}$.

In general case a particle captured to a trap is affect by an electromagnetic field, 
in particular, an external magnetic field $\overrightarrow{B}$. The energy of interaction 
of the particle with the external field should be added to $\hat{U}$. As a result, if
in a gas cell there is a magnetic field $\overrightarrow{B}$, then the potential
energy of a particle in a trap is
\[
\hat{U}=-\frac{2 \pi \hbar^2}{M \gamma_b} \rho_b \hat{f}(0)-\overrightarrow{\mu_{n}}\cdot\overrightarrow{B},
\]
where $\overrightarrow{\mu_{n}}$ is the magnetic moment of a proton (nucleus) captured in a trap.
Typically, the proton (nucleus) occur appears in the trap as a constituent part of an atom.
In this case to describe the behavior of proton (nucleus) spin one should use Shr\"{o}dinger 
equation for spin wavefunction 
\[
i\hbar \frac{\partial \Psi }{\partial t}=(H_{A}+\widehat{U})\Psi, 
\]
where $H_{A}$ is the spin Hamiltonian of the atom, 
which describes hyperfine interaction of the proton and electron interaction in the atom,
\[
\hat U =-\frac{2 \pi \hbar^2}{M \gamma_b} \rho_b \hat{f}(0)-\overrightarrow{\mu_{n}}\cdot\overrightarrow{B}
-\overrightarrow{\mu_{e}} \cdot \overrightarrow{B},
\]
$\overrightarrow{\mu_{e}}$ is the magnetic moment of an electron.

\section{Deuteron spin rotation and oscillations}

Suppose a deuteron is placed to a gas trap. 
Deuteron spin is equal to 1. This makes the picture of spin behavior more rich.
Potential energy of interaction of the deuteron with a particle beam passing 
through the trap and external fields can be expressed as follows:

\begin{equation}
\hat{U}=-\frac{2 \pi \hbar^2}{M \gamma_b} \rho_b \hat{f}_d(0)-
\overrightarrow{\mu}_d \cdot \overrightarrow{B}
\label{U}
\end{equation}
where $\overrightarrow{\mu}_d$ is the operator of deuteron magnetic moment, 
$M$ is the mass of deutron, $\hat{f}_d(0)$ is the amplitude of elastic coherent 
scattering of the deuteron by beam particles .

In contrast to the particles with spin $1/2$ rotation and oscillation of the deuteron spin
appear even while a deuteron passes through unpolarized medium.
Therefore, let us consider in details the case when a beam of nonpolarized particles
passes through a trap. 
Let us also suppose that there is no external magnetic field.
It is
important that even for unpolarized medium the amplitude of zero-angle
scattering $\hat{f}_d(0)$ is a function of the
particle spin operator and can be written as

\begin{equation}
\hat{f}_d\left( 0\right)
=d+d_{1}S_{z}^{2} 
\label{f_hat}
\end{equation}

The quantization axis $z$ is directed along $\vec{n}=\frac{\overrightarrow{k}%
}{k}$. Consider a specific case of strong interactions invariant under space
and time reflections. For this reason, the terms containing odd powers of $S$
are neglected. 

Correspondingly, the contribution to the effective potential energy, 
which is caused by the interaction of a deuteron with a beam of passing particles, can 
be expressed as
\begin{equation}
\hat{U}_{eff}=-\frac{2\pi \hbar^2 }{M \gamma_{b}} \rho_{b} \left(
d+d_{1}S_{z}^{2}\right)  \label{N_hat}
\end{equation}

From eq. (\ref{N_hat}) one can draw an important conclusion about
$U_{eff}$ being dependent on the spin orientation with respect to the
momentum direction. Suppose $m$ denotes the magnetic quantum number, then for a
particle in the state, which is an eigenstate of the spin projection operator $S_{z}$, 
one can express $U_{eff}$ as follows:
\begin{equation}
U_{eff}\left( m\right) =-\frac{2\pi \hbar^2 }{M \gamma_{b}} \rho_{b}
\left(d+d_{1}m^{2}\right)  \label{N_m}
\end{equation}

According to (\ref{N_m}), the particle states with the quantum numbers $m$
and $-m$ have the same $U_{eff}$. 
As one can see, the picture of splitting of deuteron energy levels  
under the action of the passing beam 
coincides with that of splitting of atom energy levels in an electric field due to the 
Stark effect.
Thus one can guess that a deuteron in a medium undergoes the action of some effective
quasielectric nuclear field, caused by nuclear interactions of the deuteron with 
the passing proton (nucleus).

The spin wave function of a deuteron captured to a trap at the time moment $t=0$ 
can be represented as a superposition of
basis spin wave functions $\chi _{m}$, which are the eigenfunctions of the
operators $\hat{S}^{2}$ and $\hat{S}_{z},$ 
$\hat{S}_{z}\chi_{m}=m\chi _{m}$:
\begin{equation}
\psi =\sum_{m=\pm 1,0}a^{m}\chi _{m}.
\end{equation}
The wave function of a
particle at the time moment $t$ can be expressed as:
\begin{equation}
\Psi(t) =\left\{ 
\pmatrix{
a^{1}(t) \cr 
a^{0}(t) \cr 
a^{-1}(t)
}
\right\} =\left\{ 
\pmatrix{
a\,e^{i\delta _{1}}e^{-\frac{i}{\hbar}U_{1}t} \cr 
b\,e^{i\delta _{0}}e^{-\frac{i}{\hbar}U_{0}t} \cr 
c\,e^{i\delta _{-1}}e^{-\frac{i}{\hbar}U_{-1}t}
}
\right\} =\left\{ 
\pmatrix{
a\,e^{i\delta _{1}}e^{-\frac{i}{\hbar}U_{1}t} \cr
b\,e^{i\delta _{0}}e^{-\frac{i}{\hbar}U_{0}t} \cr 
c\,e^{i\delta _{-1}}e^{-\frac{i}{\hbar}U_{1}t}
}
\right\},
\end{equation}
at the initial time moment $a^{1}(0)=a\,e^{i\delta _{1}},~a^{0}(0)=b\,e^{i\delta _{0}},~a^{-1}(0)=c\,e^{i\delta _{-1}},~
\delta _{m}$ are the initial phases.
It should be reminded that $U_{1}=U_{-1}$.

Let us choose coordinate system in which plane $\left( xz\right) $
coincides with that formed by the vectors $\langle \overrightarrow{S}\rangle $
(\bigskip $<\vec{S}\mathbf{>=}\frac{<{\psi }\mathbf{|}\vec{S}\mathbf{|}{\psi 
}>}{\mid \psi \mid ^{2}})$ and $\overrightarrow{n}=\frac{\overrightarrow{k}}{k}$ 
in the time moment $t=0$. 
In this case $\delta _{1}-\delta _{0}=\delta _{-1}-\delta _{0}=0$
and at $t=0$ the components  $langle S_{x}\rangle \neq 0,$ $\langle
S_{y}\rangle =0$ and $\langle S_{z}\rangle \neq 0$.

As a result we obtain:

\begin{eqnarray}
&<&{S}_{x}>=\sqrt{2}
e^{-\frac{1}{2}\rho (\sigma _{0}+\sigma _{1})ct}
b(a+c)
\cos [\frac{2\pi \hbar \rho_{b}}{M \gamma_{b}}{Re}d_{1}\,t]
/|\psi |^{2},  \nonumber \\
&<&{S}_{y}>=-\sqrt{2}
e^{-\frac{1}{2}\rho (\sigma _{0}+\sigma _{1})ct}
b(a-c)
\sin [\frac{2\pi \hbar \rho_{b}}{M \gamma_{b}}{Re}d_{1}t]
/|\psi |^{2},  \label{S_}
\\
&<&{S}_{z}>=
e^{-\rho \sigma _{1}ct\,}
(a^{2}-c^{2})
/|\psi |^{2}, 
\nonumber
\end{eqnarray}
$\sigma _{0}=\frac{4 \pi}{k}Im~{d}$ is the total cross-section of
scattering of a deuteron in the state with $m=0$ by a beam particle;
$\sigma _{1}=\frac{4 \pi}{k}Im~{(d+d_1)}$ is the total cross-section of
scattering of a deuteron in the state with $m=1$ by a beam particle.

Particle with spin 1 also possesses tensor polarization i.e. tensor of
rank two {\ $\hat{Q}_{ij}=3/2(\hat{S}_{i}\hat{S}_{j}+\hat{S}%
_{j}\hat{S}_{i}-4/3\delta _{ij})$ }.

For it we can obtain

\begin{eqnarray}
&<&{Q}_{xx}>=
\left\{ 
- \frac{1}{2} \left[ a^{2}+c^{2}\right] 
\,e^{-\rho \sigma_{1}\,ct}+
{b}^{2}\,e^{-\rho \sigma _{0}\,ct}+
3ac\, e^{-\rho \sigma_{1}\,ct}
\right\} 
/|\psi|^{2}  \nonumber \\
&<&{Q}_{yy}>=
\left\{ -  \frac{1}{2} \left[ a^{2}+c^{2}\right] \,
e^{-\rho \sigma_{1}\,ct}+
{b}^{2}\,e^{-\rho \sigma _{0}\,ct}-
3ac\, e^{-\rho \sigma_{1}\,ct}
 \right\} /|\psi
|^{2}  \nonumber \\
&<&{Q}_{zz}>=
\left\{ \left[ a^{2}+c^{2}\right]\,
e^{-\rho \sigma_{1}\,ct}-
2{b}^{2}\,e^{-\rho \sigma _{0}\,ct}\right\} /|\psi |^{2}~,
\label{quadr} \\
&<&{Q}_{xy}>=0,  \nonumber \\
&<&{Q}_{xz}>=\frac{3}{\sqrt{2}}
e^{-\frac{1}{2}\rho (\sigma _{0}+\sigma_{1})ct}
b(a-c)
\cos [\frac{2\pi \hbar \rho_{b}}{M \gamma_{b}}{Re}d_{1}t]
/|\psi|^{2},  \nonumber \\
&<&{Q}_{yz}>=-~\frac{3}{\sqrt{2}}
e^{-\frac{1}{2}\rho (\sigma _{0}+\sigma_{1})ct}
b(a+c)
\sin [\frac{2\pi \hbar \rho_{b}}{M \gamma_{b}}{Re}d_{1}t]
/|\psi|^{2}~,  \nonumber
\end{eqnarray}

\noindent where $|\psi |^{2}=(a^{2}+c^{2})\,
e^{-\rho \sigma_{1}\,ct}+{}b^{2}\,
e^{-\rho \sigma _{0}\,ct}$.

According to (\ref{S_},\ref{quadr}) the rotation appears if the angle
between the polarization vector and momentum of a particle differs from $\frac{\pi 
}{2}$. At this for the acute angle between polarization vector and momentum the
sign of rotation is opposite than that for the obtuse angles.

If spin is orthogonal to momentum then $(a=c)$ and particle spin (tensor of
quadrupolarization) oscillate (does not rotate)

\begin{eqnarray}
&<&{S}_{x}>=\sqrt{2}
e^{-\frac{1}{2}\rho (\sigma _{0}+\sigma _{1})ct}
2ab\cos [\frac{2\pi \hbar \rho_{b}}{M \gamma_{b}}{Re}d_{1}t]/|\psi |^{2}, \nonumber  \\
&<&{S}_{y}>=0,   \\
&<&{S}_{z}>=0,  \nonumber
\end{eqnarray}

And tensor of quadrupolarization:

\begin{eqnarray}
&<&{Q}_{xx}>=\biggl\{2 a^{2}\,
e^{-\rho \sigma _{1}\,ct}+
b^{2}\,e^{-\rho \sigma _{0}\,z}\biggr\}/|\psi |^{2}~,  
\nonumber \\
&<&{Q}_{yy}>=\biggl\{-4{a}^{2}\,
e^{-\rho \sigma _{1}\,ct}+
{b}^{2}\,
e^{-\rho \sigma _{0}\,ct}
\biggr\}/|\psi |^{2}~,  \nonumber \\
&<&{Q}_{zz}>=\biggl\{ 2{a}^{2}\,
e^{-\rho \sigma _{1}\,ct}-
2\,{b}^{2}\,
e^{-\rho \sigma _{0}\,ct}\biggr\}/|\psi |^{2}~,  \label{quad} \\
&<&{Q}_{xy}>=0,  \nonumber \\
&<&{Q}_{xz}>=0,  \nonumber \\
&<&{Q}_{yz}>=
\biggl\{-\frac{3}{\sqrt{2}}
{e}^{-\left( \sigma _{0}+\sigma_{1}\right) ct}\,{2}ab\,
\sin [\frac{2\pi \hbar \rho_{b}}{M \gamma_{b}}{Re}d_{1}t]\biggr\}%
/|\psi |^{2}~~,  \nonumber
\end{eqnarray}

\noindent where $|\psi |^{2}=2{a}^{2}\,e^{-\rho \sigma _{1}\,ct}+{b}^{2}\,e^{-\rho \sigma
_{0}\,ct}$.

Thus, according to the above analysis spins of polarized atoms captured 
to the gas cell
rotate and oscillate under the action of the beam of nonpolarized high energy particles. 
From 
(32-35) 
it follows that oscillations damp with time and
when $t \gg \frac{1}{\frac{1}{2} \rho (\sigma_0 + \sigma_1) c}$ only nonoscillating
components $<S_z>,~<Q_{xx}>,~<Q_{yy}>,~<Q_{zz}>$ remain.

Suppose now that at the initial time moment atoms in the a cell are nonpolarized. 
In this case 
(due to $\sigma_1 \ne \sigma_0$) one can conclude that after passing  the time 
$t=\rho c (\sigma_1-\sigma_0)$
deuterons will become aligned under the 
action of nonpolarized particles  i.e. they obtain $<Q_{zz}>$ 
different from zero and $<Q_{xx}>=<Q_{yy}>=-\frac{1}{2}<Q_{zz}>$ (it should be reminded that
there is the relation $<Q_{xx}>+<Q_{yy}>+<Q_{zz}>=0$ for tensor $Q_{ik}$). At this the
average value of spin of atoms in the cell 
becomes equal to zero ($<\overrightarrow S>=0$).

\section{Experimental possibilities to investigate spin rotation and oscillation of a
deuteron blowing by a nonpolarized beam rotating in a storage ring}

It should be mentioned that if deuteron in a trap exists as deuteron atom then
(\ref{U}) can be written as
\begin{equation}
\hat{U}=-\frac{2\pi {\hbar}^2 }{m \gamma} \rho {\hat{f}}_d \left( 0\right)-
\overrightarrow{{\mu}_d}\cdot \overrightarrow{B}-\overrightarrow{{\mu}_e}\cdot \overrightarrow{B}
\label{U2}
\end{equation}
where ${\hat{f}}_d$ is the amplitude of scattering of the deuteron atom 
by the beam particles , ${\mu}_e$ is the 
operator of magnetic moment of electron in deuteron atom.
In this case, to find the angle of the atom spin rotation, we should consider Hamiltonian 
${\hat{H}}={\hat{H}}_d+{\hat{U}}$, where ${\hat{H}}_d$ is the spin Hamiltonian, describing 
hyperfine interaction of 
deuteron and electron in a deuteron atom.

Using $\hat{U}$ one can find the behavior of
spin wave function and other spin characteristics of a particle (atom) at 
any certain time moment with the help of Shr\"{o}dinger equation. 
\begin{equation}
i \hbar \frac{\partial \psi }{\partial t}={\hat{H}}_A \psi.
\label{Shr_eq}
\end{equation}
%


According to the above analysis, 
to evaluate the considering effect it is necessary to 
estimate $d_1$. For this purpose let us consider the amplitude of deuteron scattering by
a proton depending on spin orientation with respect to deuteron momentum.

For fast deuterons the scattering amplitude can be found
in the eikonal approximation \cite{Cryz,hand}. According to \cite{hand}, the amplitude 
of zero-angle coherent scattering in this approximation can be written as follows:
\begin{eqnarray}
f(0)=\frac{k}{2\pi~i}\int \left( e^{i\chi _{D}\left( \overrightarrow{b},\overrightarrow{r}%
\right) }-1\right) d^{2}b\left| \varphi \left( \overrightarrow{r}\right)
\right| ^{2}d^{3}r
\label{amp}
\end{eqnarray}
where $k$ is the deuteron wavenumber, $\overrightarrow{b}$ is the impact parameter,
i.e. the distance between the deuteron and the proton centres of gravity in the normal 
plane; $\varphi \left( \overrightarrow{r}\right)$ is the wavefunction of the deuteron 
in the ground state;
$\left| \varphi \left( \overrightarrow{r}\right)\right| ^{2}$ is the probability to find 
proton and neutron (in the deuteron) at the distance $\overrightarrow{r}$ apart. 
The phase shift due to the 
deuteron scattering by a proton is
\begin{equation}
\chi _{D}=-\frac{1}{\hbar v}
\int_{-\infty }^{+\infty }V_{D}\left( \overrightarrow{b},z^{^{\prime }},%
\overrightarrow{r}_{\perp }\right) dz^{^{\prime }}
\end{equation}
$\overrightarrow{r}_{\perp}$ is the $\overrightarrow{r}$ component perpendicular 
to the momentum of incident deuteron, $v$ is the deuteron velocity. 
The phase shift $\chi _{D}=\chi _{1}+\chi _{2}$, where $\chi_{1}$ and $\chi _{2}$ are the phase 
shifts caused by proton-proton and neutron-proton interactions, respectively.

For the polarized deuteron under consideration the probability  
$\left| \varphi \left( \overrightarrow{r}\right)\right| ^{2}$ is distinguished 
for different spin states of deuteron. Thus for states with magnetic 
quantum number $m=\pm 1$, the probability is 
$\left| \varphi_{\pm 1} \left( \overrightarrow{r}\right)\right| ^{2}$,
whereas for $m=0$, it is 
$\left| \varphi_{0} \left( \overrightarrow{r}\right)\right| ^{2}$.
Owing to the additivity of phase shifts, equation (\ref{amp}) can be rewritten as
\begin{equation}
f\left( 0\right) =\frac{k}{\pi }\int 
\left\{
t_{1}
\left( 
\overrightarrow{b}-%
\frac{\overrightarrow{r}_{\perp }}{2}
\right) 
+t_{2}
\left( 
\overrightarrow{b}+%
\frac{\overrightarrow{r}_{\perp }}{2}
\right) 
+ 2it_{1}
\left( 
\overrightarrow{b
}-\frac{\overrightarrow{r}_{\perp }}{2}
\right) 
t_{2}
\left( 
\overrightarrow{b}%
+\frac{\overrightarrow{r}_{\perp }}{2}
\right)
\right\} 
\left| \varphi 
\left( 
\overrightarrow{r}
\right) 
\right| ^{2}
d^{2}bd^{3}r
\label{42}
\end{equation}
where
\[
t_{1(2)}=\frac{e^{i\chi _{1\left( 2\right) }}-1}{2i}.
\]

Attention should be given 
to the fact that the latter expression is 
valid if one neglets spin dependence of nuclear forces between the colliding
proton (neutron) and proton. When this dependence is taken into account, the
phase shift $\chi_{D}$ is an operator acting in the spin space of colliding 
particles and in the general case the expansion (\ref{42}) is not valid.
However, to estimate the magnitude of the effect of deuteron spin 
oscillations, spin dependence of nuclear forces may be negleted 
(spin dependent contribution was considered in 
\cite{spin}). Let us also omit terms caused by coulomb electromagnetic interaction.
From (\ref{42}) it follows
\begin{equation}
f(0)=f_{1}(0)+f_{2}(0)+ \frac{2ik}{\pi} 
\int t_{1}\left( \overrightarrow{b}-%
\frac{\overrightarrow{r}_{\perp }}{2}\right)t_{2}\left( \overrightarrow{b}+%
\frac{\overrightarrow{r}_{\perp }}{2}\right)\left| \varphi \left( 
\overrightarrow{r}_{\perp},z\right) \right| ^{2}d^{2}bd^{2}r_{\perp}dz
\label{integral}
\end{equation}
where 
\[
f_{1(2)}(0)=\frac{k}{\pi} \int t_{1(2)}(\overrightarrow{\xi})d^{2}\xi=
\frac{m_D}{m_{1(2)}}~f_{p(n)}(0)
\]
and $f_{p(n)}(0)$ is the amplitude of the proton (neutron)-proton zero-angle
elastic coherent scattering.

The expression (\ref{integral}) can be rewritten as
\begin{equation}
f(0)=f_{1}(0)+f_{2}(0)+
\frac{2ik}{\pi}\int t_{1}(\overrightarrow{\xi})~
t_{2}(\overrightarrow{\eta}) 
\left| \varphi \left(\overrightarrow{\xi}-\overrightarrow{\eta},z\right) 
\right| ^{2}~d^{2}\xi~d^{2}\eta~dz
\label{27}
\end{equation}

Then from (\ref{27})
\begin{eqnarray}
&Re&~f(0)=Re~f_{1}(0)+Re~f_{2}(0)
-\frac{2k}{\pi}Im \int t_1(\overrightarrow{\xi})
t_{2}(\overrightarrow{\eta})\left| 
\varphi \left(\overrightarrow{\xi}-\overrightarrow{\eta},z\right) 
\right| ^{2}~d^{2}\xi~d^{2}\eta~dz \\ \nonumber
&Im&~f(0)=Im~f_{1}(0)+Im~f_{2}(0)+
+\frac{2k}{\pi}Re \int t_1(\overrightarrow{\xi})
t_{2}(\overrightarrow{\eta})\left| \varphi \left(\overrightarrow{\xi}-\overrightarrow{\eta},z\right) 
\right| ^{2}~d^{2}\xi~d^{2}\eta~dz \nonumber
\label{28}
\end{eqnarray}


In accodance with (\ref{S_},\ref{quadr}) spin oscillation period is determined by the difference of 
the amplitudes 
$Re~f(m=\pm1)$ and $Re~f(m=0)$. From (\ref{42}) it follows that
\begin{eqnarray}
Re~d_1=-\frac{2k}{\pi}Im \int
t_1(\overrightarrow{\xi})
t_{2}(\overrightarrow{\eta})\left[ 
\varphi_{\pm 1}^{+} \left(\overrightarrow{\xi}-\overrightarrow{\eta},z\right) 
\varphi_{\pm 1} \left(\overrightarrow{\xi}-\overrightarrow{\eta},z\right)- 
\varphi_{0}^{+} \left(\overrightarrow{\xi}-\overrightarrow{\eta},z\right)
\varphi_{0} \left(\overrightarrow{\xi}-\overrightarrow{\eta},z\right)
\right]~d^{2}\xi~d^{2}\eta~dz \\ \nonumber
Im~d_1=\frac{2k}{\pi}Re \int
t_1(\overrightarrow{\xi})
t_{2}(\overrightarrow{\eta})\left[ 
\varphi_{\pm 1}^{+} \left(\overrightarrow{\xi}-\overrightarrow{\eta},z\right) 
\varphi_{\pm 1} \left(\overrightarrow{\xi}-\overrightarrow{\eta},z\right)- 
\varphi_{0}^{+} \left(\overrightarrow{\xi}-\overrightarrow{\eta},z\right)
\varphi_{0} \left(\overrightarrow{\xi}-\overrightarrow{\eta},z\right)
\right]~d^{2}\xi~d^{2}\eta~dz 
\label{d1}
\end{eqnarray}

It is well known that the characteristic radius of the deuteron is large
comparing with the range of nuclear forces. For this reason, when integrating,
the functions $t_1$ and $t_2$ act on $\varphi$ as $\delta$-function. Then

\begin{eqnarray}
Re~d_1=-\frac{4k}{\pi}Im {f_1(0)~f_{2}(0)}
\int_{0}^{\infty} 
\left[ 
\varphi_{\pm 1}^{+} \left(0,z\right) 
\varphi_{\pm 1} \left(0,z\right)- 
\varphi_{0}^{+} \left(0,z\right)
\varphi_{0} \left(0,z\right)
\right]~dz \\ \nonumber
Im~d_1=\frac{4k}{\pi}Re {f_1(0)~f_{2}(0)}
\int_{0}^{\infty}
\left[ 
\varphi_{\pm 1}^{+} \left(0,z\right) 
\varphi_{\pm 1} \left(0,z\right)- 
\varphi_{0}^{+} \left(0,z\right)
\varphi_{0} \left(0,z\right)
\right]~dz 
\label{d1}
\end{eqnarray}

The magnitude of the spin oscillation effect is determined by difference
\[
\left[ \varphi_{\pm 1}^{+} \left(0,z\right) 
\varphi_{\pm 1} \left(0,z\right)- 
\varphi_{0}^{+} \left(0,z\right)
\varphi_{0} \left(0,z\right)\right]
\]
i.e. by the difference of distributions of nucleon density in the 
deuteron for different deuteron spin orientations.
The structure of the wavefunction $\varphi_{\pm 1}$ is well known:
\begin{equation}
\varphi_m=\frac{1}{4 \pi} 
\left\{
\frac{u(r)}{r}+\frac{1}{\sqrt 8}\frac{W(r)}{r}\hat{S}_{12}
\right\}
\chi_m
\label{phi_m}
\end{equation}
where $u(r)$ is the deuteron radial wavefunction corresponding to the S-wave;
$W(r)$ is the radial function corresponding to the D-wave; the operator 
$\hat{S}_{12}=6(\hat{\overrightarrow{S}} \overrightarrow{n}_{r})^2-2\hat{\overrightarrow{S}}^2$;
$\overrightarrow{n}_{r}=\frac{\overrightarrow{r}}{r}$; 
$\hat{\overrightarrow{S}}=\frac{1}{2}(\overrightarrow{\sigma}_1+\overrightarrow{\sigma}_2)$ 
and $\overrightarrow{\sigma}_{1(2)}$ ate the Pauli spin matrices 
describing proton(neutron) spin.

Use of (\ref{phi_m}) yields
\begin{eqnarray}
Re~d_1=-\frac{3}{2 \pi k}~Im
\left\{
f_{1}(0)f_{2}(0)
\right\} G = -\frac{\sigma}{\pi k} Im \left\{
f_{p}(0)f_{n}(0)
\right\} G          \\ \nonumber
Im~d_1=\frac{3}{2 \pi k}~Re
\left\{
f_{1}(0)f_{2}(0)
\right\} G= \frac{\sigma}{\pi k} Re \left\{
f_{p}(0)f_{n}(0)
\right\} G     
\end{eqnarray}
where $G=\int_{0}^{\infty}
\left(
\frac{1}{\sqrt 2}\frac{u(r)W(r)}{r^2}-\frac{1}{4} \frac{W^2(r)}{r^2}
\right) dr
$, $r^2=\xi^2+z^2$.
Applying the optical theorem
$Im~f=\frac{k}{4 \pi}~\sigma$, where $\sigma$ is the total scattering cross-section,
one can obtain
\begin{eqnarray}
Re~d_1=-\frac{3}{2 \pi^2}
\left(
Re~f_p(0) \sigma_{n}+Re~f_n(0) \sigma_{p}
\right) G \\
Im~d_1=\frac{3}{2 \pi k}
\left(
Re~f_1~Re~f_2-\frac{k^2}{(4 \pi)^2}\sigma_{1}\sigma_{2}
\right) G =
\left(
\frac{6}{\pi k} Re~f_p~Re~f_n-\frac{3k}{(2 \pi)^3}\sigma_{p}\sigma_{n}
\right) G
\end{eqnarray}
where $\sigma_{p(n)}$ is the total cross-section 
of the proton-proton (neutron-proton)nuclear scattering.




Now we can evaluate the phase of spin oscillation of the deuterons captured in trap,
which appears under the action of the beam of unpolarized particles rotating in the 
storage ring.

\begin{equation}
\varphi =\frac{2\pi \hbar \rho_{b} \;t}{M \gamma_{b}}{Re}d_{1}.
\label{fi}
\end{equation}
The expression (48) could be rewritten as follows:
\begin{equation}
\varphi =2\pi \frac{\hbar}{Mc}\frac{{Re}d_{1}}{\gamma_{b}}~\rho_{b}~ct=
2 \pi \lambda_{c}\frac{{Re}d_{1}}{\gamma_{b}}\frac{N(t)}{S_b},
\label{fi_1}
\end{equation}
where $\lambda_{c}=\frac{\hbar}{Mc}$ is the Compton wavelength, 
$N(t)=S_b \rho_{b}ct=N_c \nu t$ is the number 
of particles passed through the target during the time $t$, $S_b$ is the cross-section
of the particle beam, which passes through the target,  $N_c$ is the number of particles
rotating in the storage ring, $\nu$ is the frequency of their rotation in orbit.
Suppose that $N_c=10^{11}$, $\nu=10^6$ $s^{-1}$ and $\frac{{Re}d_{1}}{\gamma_{b}}$ one can
obtain $\varphi=10^{-9} \frac{t}{S_b}$. Therefore, for $S_b=10^{-6}$ $cm^2$ and $t=10^3 s$
$\varphi \sim 1 ~rad$.

%
%
%
%
Attention should be drawn to the fact that hydrogen atom in triplet state has spin 1.
%
Therefore,
all the above is relevant for a cell containing hydrogen atoms in triplet state, too.
According to \cite{6rot} study of deuteron spin-dependent scattering amplitude
allows, among other things, to investigate real part of nucleon-nucleon scattering amplitude. 

If a cell contains hydrogen atoms, then studying spin oscillation and rotation of hydrogen atom
blowing by a flow of nonpolarized protons one can reconstruct real part of proton-proton scattering
amplitude.


\section{Conclusion}

Thus, the above analysis shows that, as a result of coherent rescattering processes, a particle
in a gas cell undergoes action of effective quasimagnetic and quasielectric fields, 
formed by the flow of high energy particles. 
Picture of energy levels splitting, necessary for analysis of behavior of particle spin,
can be obtained from Shr\"{o}dinger equation for spin wave function of a particle (\ref{Shr_eq})
\begin{equation}
i \hbar \frac{\partial \psi }{\partial t}=\hat{U}_{eff} \psi,
\end{equation}

If a particle beam flying at a proton (deuteron, atom, molecule) containing in a cell 
has non-zero polarization, then
from general structure of scattering amplitude \cite{5rot} we can express $U_{eff}$ as
\begin{equation}
\hat{U}_{eff}=-\frac{2 \pi {\hbar}^2}{m \gamma} \rho \hat{f}(0)-\overrightarrow{\mu}_N\overrightarrow{B}
-\overrightarrow{\mu}_e \overrightarrow{B}
\end{equation}
where $\overrightarrow{\mu}_N$ and $\overrightarrow{\mu}_e $ are the magnetic moments of 
nuclei and electron shell of atom, respectively,
\begin{eqnarray}
\hat{f}(0)=A+A_1(\overrightarrow{S}\overrightarrow{p})
+A_{2}\left( \overrightarrow{S}\overrightarrow{n}\right) \left( 
\overrightarrow{p}\overrightarrow{n}\right) + \\
+d_{1}\left( \overrightarrow{S}%
\overrightarrow{n}\right) ^{2}+B_0 \left( \overrightarrow{S}\overrightarrow{n}%
\right) +B_{1}\overrightarrow{S}\left[ \overrightarrow{p}\times 
\overrightarrow{n}\right] + \nonumber \\
+ B_{2}\left( \overrightarrow{S}\overrightarrow{n}\right) \left( 
\overrightarrow{S}\overrightarrow{p}\right) +B_{3}\left( \overrightarrow{S}%
\overrightarrow{n}\right) ^{2}\left( \overrightarrow{p}\overrightarrow{n}%
\right) + \nonumber \\
+ B_{4}\left( \overrightarrow{p}\overrightarrow{n}\right) +
B_{5}\left( \overrightarrow{S}\left[ \overrightarrow{p}\times 
\overrightarrow{n}\right] \right) \left( \overrightarrow{S}\overrightarrow{n}%
\right) +\,.... \nonumber
\end{eqnarray}

\noindent where terms containing $A$ and $d$ are caused by strong P-,T-even interactions,
those with $B_1,~B_2,~B_3,~B_4$ - by P-odd T-even, $B_0$ - by P-odd T-odd and 
$B_5$ corresponds to P-even T-odd interactions.
It should be emphasized that $\hat{f}(0)$ is amplitude of scattering by atom as a whole, therefore 
it includes both scattering by atom nucleus and atom electrons,
$\overrightarrow{S}$ is the spin of an atom.

It should be emphasized that in discussed experiment arrangement the effect of
rotation and oscillation of spin of particles captured in a trap
can exceed those investigated for high energy particles passing through a trap.
This is caused by the following.
In the first case (when we observe spin rotation of particles captured in a trap, see 
Fig.\ref{second})
density of particles in a trap $\rho$ can be quite low 
$(\rho \sim 10^8-10^9~~ cm^{-3})$. As a result
life-time of a particle beam rotating in a storage ring does not change.
But in the second case (when we observe spin rotation of high energy particles, 
see Fig.\ref{conventional})
the density of particles in gas cell should be rather high (for example,
$\rho \sim 10^{12}~cm^{-3}$) to increase the effect value, but density growth yields to
raising of multiple scattering by gas and, as a result, reduces life time of high energy particle 
in a storage ring.
Therefore, in the first case observation time can exceed that in the 
second case and effect can grow.

However, to carry out the above experiment one need to develop a trap, in which polarized atoms could 
live for a long time and, simultaneously, which is transparent for high energy particles.


\begin{thebibliography}{99}

\bibitem{Lehar} C.Lechanoine-Lelue and F.Lehar Rev.Mod.Phys. {\bf 65} (1993) 47.

\bibitem{2rot}  V.G. Baryshevsky, Sov. J. Nucl.Phys. {\bf 38} (1983) 569; V.G.
Baryshevsky, Phys. Lett. {B} 120 (1983) 267.

\bibitem{3rot}  V.G. Baryshevsky, I.Ya.Dubovskaya, Phys. Lett. {B} {\bf 256}
(1989) 529.

\bibitem{4rot}  V.G. Baryshevsky, A.G. Shekhtman, Phys. Rev. C {\bf 53}, n.1
(1996) 267.

\bibitem{5rot}  V. G. Baryshevsky, Phys. Lett. {\bf 171A} (1992) 431. 

\bibitem{6rot}  V. G. Baryshevsky, J. Phys.G {\bf 19} (1993) 273. 

\bibitem{7rot}  V. G. Baryshevsky, K. G. Batrakov and S. Cherkas J. Phys.G
{\bf 24} (1998) 2049.

\bibitem{8rot}  V. Baryshevsky, K. Batrakov, S. Cherkas, LANL e-print
archive: hep-ph/9907464.

\bibitem{LANL2001}  V.G. Baryshevsky, LANL e-print archive: hep-ph/0109099.

\bibitem{1rot}  V.G. Baryshevsky and M.I. Podgoretsky, Zh. Eksp. Teor.
Fiz. {\bf 47} (1964) 1050.

\bibitem{9rot}  A. Abragam et al., C.R. Acad. Sci. {\bf 274} (1972) 423.

\bibitem{10rot}  M. Forte, Nuovo Cimento A {\bf 18} (1973) 727.

\bibitem{11}  A. Abragam and M. Goldman, Nuclear magnetism: order and
disorder (Oxford Univ. Press, Oxford, 1982).

\bibitem{12}  M.Lax, Rev.Mod.Phys. {\bf 23} (1951) 287.

\bibitem{Goldberger} M.L. Goldberger and K.M. Watson, Collision Theory: John Wiley, New York, 1964.

\bibitem{14}  L.D. Landau and E.M. Lifshits, Quantum mechanics: Pergamon,
New York, 1984.

\bibitem{Cryz} W. Cryz, L.C. Maximon Ann.Phys., NY {\bf 52} (1969) 59

\bibitem{hand} Handbuch der Physik {\bf 39} (Berlin: Springer, 1957) 112








\end{thebibliography}
\end{document}